\documentclass[aps,prb,twocolumn,showpacs]{revtex4}
\usepackage{graphicx}
\begin{document}
\draft

\title{Electric Field Effects on the Diffuse Scattering in PZN-8\%PT}

\author{P. M. Gehring,$^1$ K. Ohwada,$^2$ and G. Shirane$^3$}

\address{$^1$NIST Center for Neutron Research, National Institute of
  Standards and Technology, Gaithersburg, Maryland 20899-8562}

\address{$^2$Synchrotron Radiation Research Center (at SPring 8),
Japan Atomic Energy Research Institute, Sayo-gun Hyogo 679-5148,
Japan}

\address{$^3$Department of Physics, Brookhaven National Laboratory,
  Upton, New York 11973-5000}

\begin{abstract}
We report measurements of the neutron diffuse scattering from a
single crystal of the relaxor ferroelectric
PbZn$_{1/3}$Nb$_{2/3}$O$_3$ doped with 8\%\ PbTiO$_3$ (PZN-8\%PT)
for temperatures 100\,K $\le T \le $ 530\,K and electric fields
0\,kV/cm $\le E \le $ 10\,kV/cm.  The diffuse scattering is
strongly suppressed transverse to the (003) Bragg peak at 400\,K
for $E = 2$\,kV/cm applied along the [001] direction.  However, no
change in the diffuse scattering is observed transverse to (300),
even for field strengths up to 10\,kV/cm.  Thus the application of
an external electric field in the tetragonal phase does not
produce a macroscopically polarized state.  This unusual behavior
can be understood within the context of the Hirota {\it et al.}
phase-shifted polar nanoregion (PNR) model of the relaxor diffuse
scattering, since an electric field $E$ oriented parallel to [001]
would remove the shift of the nanoregions along [001] while
preserving those along the orthogonal [100] direction.
\end{abstract}


\date{\today}
\maketitle

\section{Introduction}

The shape, size, density, and orientation of the polar nanoregions
(PNR) that form below the Burns temperature $T_d$~\cite{Burns} in
the lead-oxide class of relaxors Pb(Zn$_{1/3}$Nb$_{2/3}$)O$_3$ and
Pb(Mg$_{1/3}$Nb$_{2/3}$)O$_3$ doped with PbTiO$_3$ (PZN-$x$PT and
PMN-$x$PT, respectively) are key parameters of interest to many
researchers attempting to clarify the various mechanisms
responsible for the exceptional piezoelectric properties exhibited
by these materials.~\cite{Park}  First identified through optical
measurements of the index of refraction by Burns and
Dacol,~\cite{Burns} the presence of PNR have been correlated with
the onset of neutron diffuse scattering in PMN,~\cite{Naberezhnov}
a saturation of the unit cell volume below $T_d$,~\cite{Zhao} and
the overdamping of long-wavelength phonon modes in PZN and
PZN-8\%PT.~\cite{Gehring_pzn,Gehring_pzn8pt}  More recently an
important model concerning the relationship between the PNR and
the surrounding cubic lattice was presented by Hirota {\it et
al.}~\cite{Hirota}  Based on neutron time-of-flight measurements
of the diffuse scattering in PMN at 300\,K performed by Vakhrushev
{\it et al.},~\cite{Vakhrushev} Hirota {\it et al.} noticed that
the corresponding ionic displacements could be decomposed into a
scalar shift (displacement) common to all atoms, and a pattern of
shifts that satisfies a center-of-mass condition.~\cite{Harada} It
was further shown that these center-of-mass shifts are consistent
with neutron inelastic measurements of the soft TO phonon
intensity in PMN above $T_d$, thus providing compelling evidence
that the polar nanoregions result from the condensation of a soft
mode at the Burns temperature.

Subsequent measurements on PZN-8\%PT using neutron diffraction
techniques by Ohwada {\it et al.} demonstrate that the zero-field
cooled structure at low temperature is not rhombohedral as had
been previously believed.~\cite{Ohwada}  Instead, surprising
evidence of a new phase, called phase X, was found that exhibits
an average cubic unit cell structure. These findings motivated the
high-energy x-ray study by Xu {\it et al.} on pure PZN,~\cite{Xu}
and the high-resolution neutron study by Gehring {\it et al.} on
PMN-10\%PT.~\cite{Gehring_pmn10pt} Both of these studies revealed
a low-temperature bulk phase that is not rhombohedral, but rather
consistent with the phase X observed by Ohwada {\it et al.} in
PZN-8\%PT.  It has been suggested that the phase-shifted nature of
the PNR could be responsible for stabilizing phase X at low
temperatures as the uniform shift of the PNR would present a
natural energy barrier in the system against the formation of a
uniform polar state.

To shed more light on this idea, we have performed a study of the
PNR in PZN-8\%PT by measuring the response of the neutron diffuse
scattering below the cubic-to-tetragonal phase transition
temperature $T_c \sim 470$\,K to an external electric field ${\bf
E}$.  As is well known, a classic ferroelectric system will break
into many domains below $T_c$, and the application of an electric
field can produce a nearly uniform polar state by aligning all of
the domains via domain wall motion.  In the case of the relaxor
ferroelectric PZN-8\%PT, however, our data show that for ${\bf E}$
oriented parallel to the [001] direction, only the diffuse
scattering transverse to the (003) reciprocal lattice point
decreases.  The diffuse scattering transverse to the orthogonal
(300) position is unaffected.  Thus a macroscopic uniformly
polarized state is not achieved.  The model of phase-shifted PNR
provides a possible explanation for these findings, as an electric
field along [001] would remove the PNR shift along [001] while
preserving that along [100].

\section{Experimental Details}

The neutron scattering data presented here were obtained on the
BT9 triple-axis spectrometer located at the NIST Center for
Neutron Research.  The diffuse scattering near the reciprocal
lattice points (300) and (003) was measured at a fixed neutron
energy $E_i = E_f = $14.7\,meV ($\lambda = 2.36$\,\AA) using the
(002) reflection of highly-oriented crystals of pyrolytic graphite
(HOPG) as monochromator and analyzer. Horizontal beam collimations
were 40$'$-46$'$-S-40$'$-80$'$ (S = sample).

The high-quality single crystal of PZN-8\%PT used in this study
was grown by the flux solution method \cite{Mulvihill} at the
Pennsylvania State University. It is a sister crystal to the one
used by Ohwada {\it et al.} in an earlier study, some results from
which are presented in this paper.~\cite{Ohwada}  The crystal
weighs 1.1\,gm (0.15\,cm$^3$), and is a rectangular block with
{100} faces and dimensions $7.2 \times 6.9 \times 3.0 $\,mm$^3$.
Gold electrodes were plated onto the two largest surfaces of the
crystal to which thin gauge copper wires were attached using a
fired-on silver conductive adhesive epoxy cured at 100°C for 2
hours.  In this geometry the field is always applied along a cubic
[001] axis. These wires were twisted and soldered onto leads
connected to a high voltage power supply. Tests were performed
both before and after the experiment to verify that the voltage
set by the power supply appeared across the sample.  The
difference between the set and measured voltage drop was always
less than 1\,\%.

The crystal was mounted onto an electrically insulating boron
nitride post using boron nitride paste to provide a strain-free
environment for the sample.  The crystal [010] axis was oriented
vertically, giving access to reflections of the form $(h0l)$. The
sample holder assembly was then mounted inside the vacuum space of
a high-temperature closed-cycle $^3$He refrigerator, which was
subsequently positioned and fixed onto the goniometer of the BT9
spectrometer.

\section{Diffuse Scattering}

The PZN-8\%PT phase $E$-$T$ diagram has been mapped out by Ohwada
{\it et al.}, and is shown in Fig.~1.~\cite{Ohwada}  The data
presented in the top panel were taken while cooling in a constant
electric field, whereas those shown in the bottom panel were taken
with increasing field at fixed temperature after first cooling in
zero field.  The electric field acts to stabilize the tetragonal
phase as can be seen from the slope of the tetragonal (T) to cubic
(C) phase boundary in Fig.~1(a), where the transition temperature
increases with field.  The transition temperatures, represented by
the circles in Fig.~1, were determined from measurements of the
lattice constants (Bragg peaks).  But while this relaxor compound
clearly exhibits a tetragonally-distorted structure, the precise
nature of the polar order is still unknown.

%
%
\begin{figure}
\includegraphics[width=3.0in]{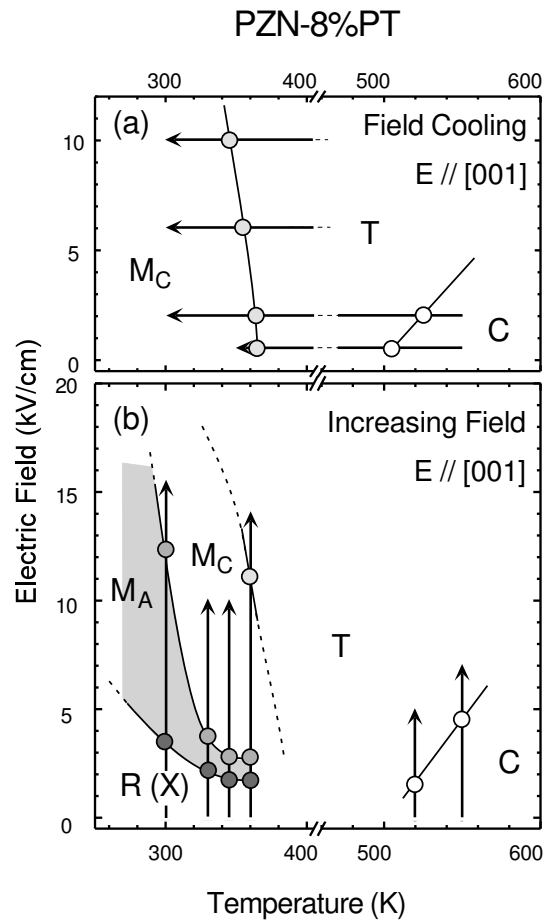}
\caption{\label{fig:1}The $E-T$ phase diagram of PZN-8\%PT
determined by Ohwada {\it et al.}~\protect\cite{Ohwada}  The
notations C, T, M, and R refer to cubic, tetragonal, monoclinic,
and rhombohedral phases, respectively. The arrows point in the
scan direction, while the length of the arrows correspond to the
scan range. The circles denote the transition temperatures
determined from each scan.}
\end{figure}
%
%

Important information can be obtained about the polar order in the
tetragonal phase of PZN-8\%PT through measurements of the diffuse
scattering.  Fig.~2 shows neutron scattering measurements taken by
Ohwada {\it et al.} (on a different PZN-8\%PT single crystal, but
grown using the same technique) near the (003) reciprocal lattice
position, for which the diffuse scattering structure factor is
known to be strong.~\cite{Ohwada} The intensity at ${\bf Q}$ =
(0.04,0,3) is shown as a function of temperature at $E = 0$\,kV/cm
(solid circles) and 3\,kV/cm (open circles), with the field
applied along the [001] direction. In the cubic phase at high
temperature ($T \ge 520$\,K) the diffuse scattering intensities
are equal, indicating no field dependence. However at lower
temperatures, in the tetragonal phase, the diffuse scattering is
strongly suppressed by the field while the zero-field diffuse
scattering intensity remains strong and increases with cooling.
For purposes of comparison, the integrated Bragg intensity
measured at (002) in zero-field is shown (scale on the right-hand
side of Fig.~2). The Bragg intensity increases sharply below
500\,K, which is consistent with the slope of the T-C phase
boundary in Fig.~1.

%
%
\begin{figure}
\includegraphics[width=3.0in]{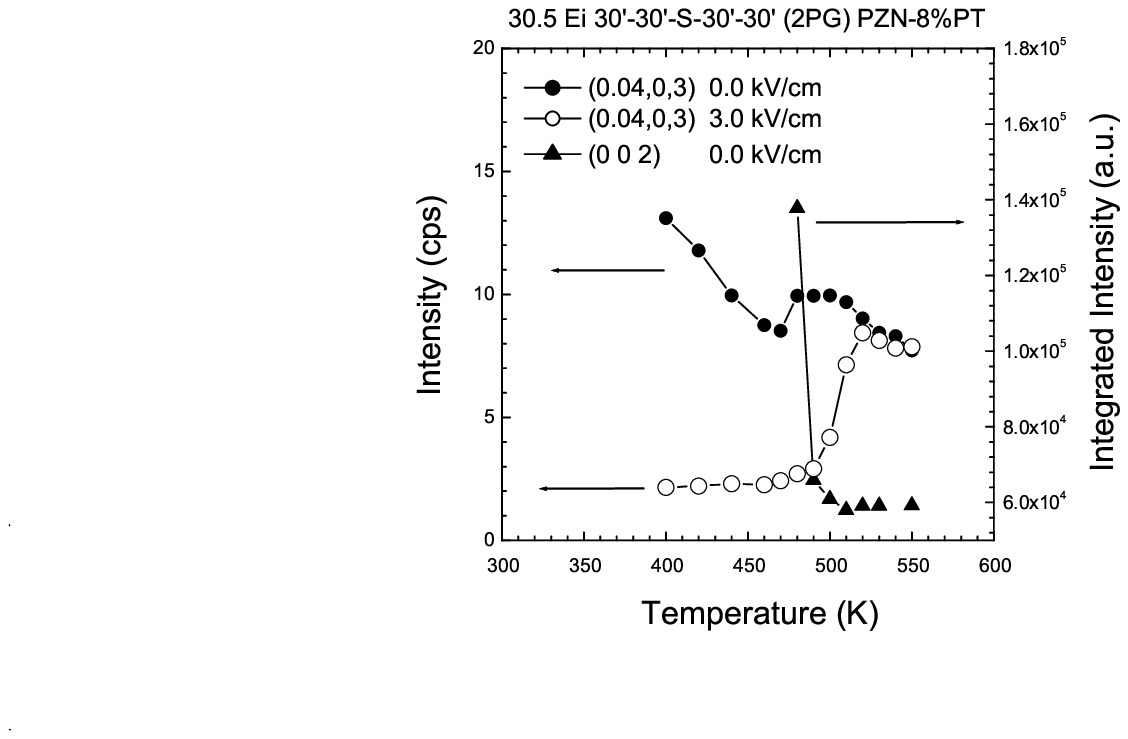}
\caption{\label{fig:2} Temperature dependence of the neutron
diffuse scattering at $\vec{Q}$ = (0.04,0,3) in zero field (solid
circles) and $E$ = 3kV/cm // [001] (open circles).  The zero-field
Bragg intensity measured at (002) is shown as a function of
temperature by the solid triangles.  These data were taken by
Ohwada {\it et al.} on a different single crystal of PZN-8\%PT.}
\end{figure}
%
%

The data in Fig.~2 were obtained at a fixed ${\bf Q}$ value at two
different field values.  Information about the geometry of the
diffuse scattering is presented in Fig.~3, which compares the
diffuse scattering intensity measured at several different
reciprocal lattice points at a fixed field strength of 2\,kV/cm
oriented along the [001] axis.  The intensities measured around
the (003) position (open circles) decrease upon cooling in field
in agreement with the data of Fig.~2.  However the striking
behavior revealed in this figure is the that the diffuse
intensities measured around the orthogonal (300) position remain
high, much like the zero-field data shown in Fig.~2.  The vertical
arrows in this figure indicate the phase transition temperatures
$T_{M-T}$ and $T_{T-C}$ between the monoclinic and tetragonal, and
tetragonal and cubic phases, respectively.

%
%
\begin{figure}
\includegraphics[width=3.0in]{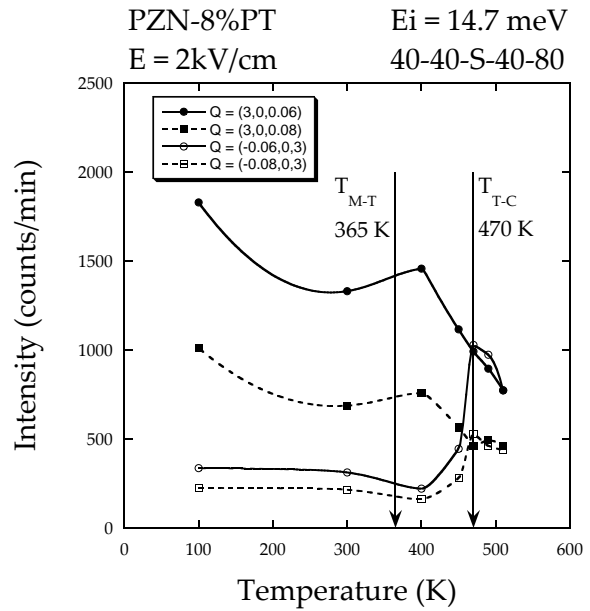}
\caption{\label{fig:3} Comparison of the diffuse scattering
intensities measured near (300) (solid symbols) and (003) (open
symbols) as a function of temperature in a field of 2kV/cm applied
along [001]. }
\end{figure}
%
%

To clarify this behavior wide scans were made of the scattered
neutron intensity as a function of the reduced momentum transfer
$q$ transverse to both the (300) and (003) Bragg positions as a
function of field and temperature.  The resulting data are plotted
on a log-intensity scale in Fig.~4.  The top two panels display
transverse $q$-scans in zero field, and the bottom two panels show
the same scans taken after cooling the sample in a 2\,kV/cm field.
In the cubic phase at 500\,K we observe diffuse scattering of
roughly equal intensity in zero field along directions transverse
to both ${\bf Q}$ = (300) and (003).  This scattering does not
change when an external electric field of 2\,kV/cm is applied
along the [001] direction.  In the tetragonal phase at 400\,K the
diffuse scattering in zero field increases slightly, becoming
somewhat stronger around (003) compared to (300).  Then when the
sample is field cooled, the diffuse scattering transverse to (003)
is reduced substantially, as is suggested by the data shown in
Fig.~3.  Remarkably, however, the diffuse scattering transverse to
(300) is unaffected.  This latter result indicates that the
application of a 2\,kV/cm field is insufficient to produce a
uniform polarization in the tetragonal phase.  The dotted line in
the lower left panel of Fig.~4 indicates the measured instrumental
(Gaussian) resolution function.  Although data are not shown, the
same behavior was observed at 300\,K in the monoclinic phase of
PZN-8\%PT as in the tetragonal phase at 400\,K.

%
%
\begin{figure}
\includegraphics[width=3.0in]{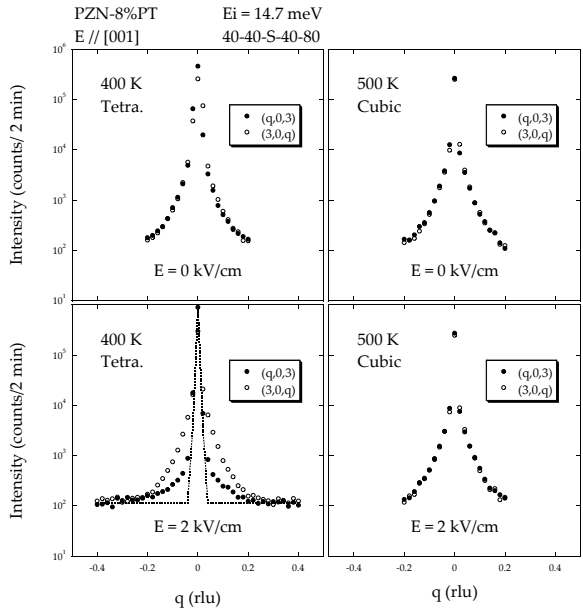}
\caption{\label{fig:4} A log-scale comparison of the diffuse
scattering intensities measured transverse to the (300) (open
circles) and (003) (solid circles) reciprocal lattice positions at
400\,K (tetragonal phase) and 500\,K (cubic phase).  The top
panels show data in zero field, and the bottom panels show data in
an electric field of 2\,kV/cm (bottom panels) applied along [001].
The dashed line in the lower left panel represents the
instrumental transverse $q$-resolution. }
\end{figure}
%
%

An effort was made to determine the field strength $E_c$ at which
the diffuse scattering transverse to (300) begins to decrease.
Fig.~5 shows the diffuse scattering intensity measured at ${\bf
Q}$ = (3,0,-0.06) at 450\,K, still in the tetragonal phase, from
zero to 10\,kV/cm.  All of these data were taken as a function of
increasing electric field strength after first field cooling from
520~K in a field of 2\,kV/cm except for the zero field data point,
which was obtained after zero-field cooling. The solid and open
circles are successive measurements at the same temperature and
${\bf Q}$. The solid square shows the diffuse scattering intensity
at 450\,K at $E$ = 2\,kV/cm.  These data clearly reveal the
inability of even a strong external electric field to diminish the
diffuse scattering orthogonal to the field direction.  Hence the
underlying mechanism responsible for the diffuse scattering in
PZN-8\%PT must have an energy barrier that is sufficiently large
in order to stabilize it against external fields of at least
10\,kV/cm.

%
%
\begin{figure}
\includegraphics[width=3.0in]{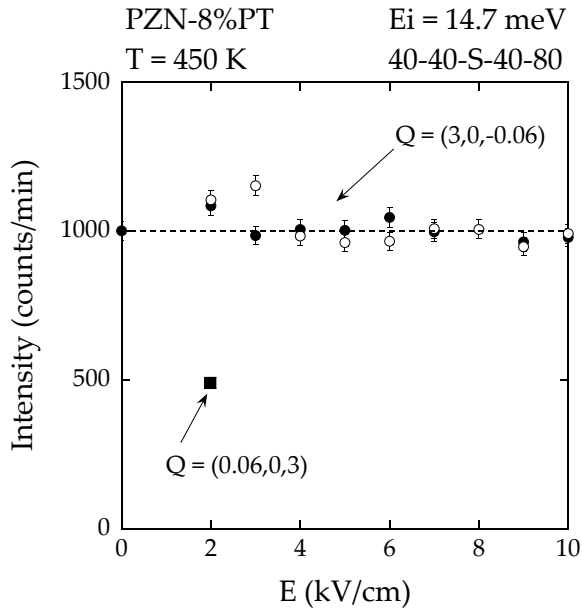}
\caption{\label{fig:5} Plot of the diffuse scattering intensity
measured in the tetragonal phase at (3,0,-0.06) at 450\,K as a
function of increasing field strength applied along the [001]
direction.  The open and solid circles represent independent
measurements.  The solid square corresponds to the diffuse
scattering intensity at 2\,kV/cm at (0.06,0,3) for comparison.}
\end{figure}
%
%

\section{Discussion}

The results on PZN-8\%PT presented here are significant not
because the diffuse scattering measured transverse to (003) is
suppressed by an external electric field, but rather because the
diffuse scattering measured transverse to (300) is \emph{not}.
Independent of any model, our data provide unambiguous evidence
that the field-induced polar state in this important relaxor
material is not macroscopically uniform.  This unusual finding
must therefore be explained by any model that purports to address
the fundamental properties of these lead-oxide relaxor compounds.
While not unique, we believe that the concept of the phase-shifted
polar nanoregions provides a natural explanation for the observed
field and ${\bf Q}$-dependence of the diffuse scattering in
PZN8\%PT.  This phase shift, or uniform displacement, of the PNR
occurs with respect to the surrounding cubic lattice at the Burns
temperature $T_d$ where the PNR condense from the soft TO phonon
mode (see the top panel in Fig.~6).  This soft TO mode is in fact
coupled to the TA mode as first shown by Naberezhnov {\it et al.},
and later studied by Wakimoto {\it et
al}.~\cite{Naberezhnov,Wakimoto1}  The uniform phase shift is
believed to result from the fact that the soft TO mode is actually
a soft-\emph{coupled} mode that contains a significant transverse
acoustic component, an idea originally proposed by Yamada and
Takakura.~\cite{Yamada}

The results of our study can be explained quite simply on the
basis of the shifted nature of the PNR, as illustrated by the
three-panel schematic diagram shown in Fig.~6.  In the top panel,
the system is in the cubic phase, below the Burns temperature. The
white squares represent the non-polar regions of the lattice (note
that these are not individual unit cells), while the PNR are shown
as the red hatched regions (corresponding to something of the
order of 10 unit cells), which are slightly displaced in different
directions from the cubic lattice.  The arrows represent the
direction and size of the uniform shift, which occurs in the
direction of the PNR polarization.  In the tetragonal phase ($T <
T_c$) the lattice develops its own polarization as indicated by
the red dotted shading, while the uniform shift of the PNR remains
intact (see the upper left panel of Fig.~4). Under the application
of an external electric field, the shifts antiparallel to ${\bf
E}$ are removed, thereby reducing the local disorder, and in turn
suppressing the diffuse scattering.  (As it is unclear how much of
the diffuse scattering transverse to the (003) reciprocal lattice
position remains (see the lower left-hand panel of Fig.~4), we can
only speculate that parallel shifts may be unaffected.)  But the
orthogonal PNR shifts are unaffected, and hence the corresponding
diffuse scattering is also unaffected.  Thus a macroscopically
ordered polar phase is not achieved in this system, even under
applied electric field strengths up to 10\,kV/cm.

Interesting high-$q$ resolution studies in zero field have now
been performed by Gehring {\it et al.} on PMN-10\%PT, and Xu {\it
et al.} on PMN-20\%PT and
PMN-27\%PT.~\cite{Gehring_pmn10pt,Xu_pmn} These studies have
examined the low-temperature phase in the PMN-$x$PT system, and
document the existence of phase X up to PbTiO$_3$ concentrations
as high as 20\%.  The 27\% sample finally exhibits the expected
rhombohedral phase.  The stability of this new phase X against the
formation of a global rhombohedral polar phase at temperatures
below $T_c$ can also be understood in terms of the phase-shifted
PNR model.  The latter study by Xu {\it et al.} in particular
demonstrates that phase X is both undistorted and rhomohedrally
polarized.

%
%
\begin{figure}
\includegraphics[width=3.0in]{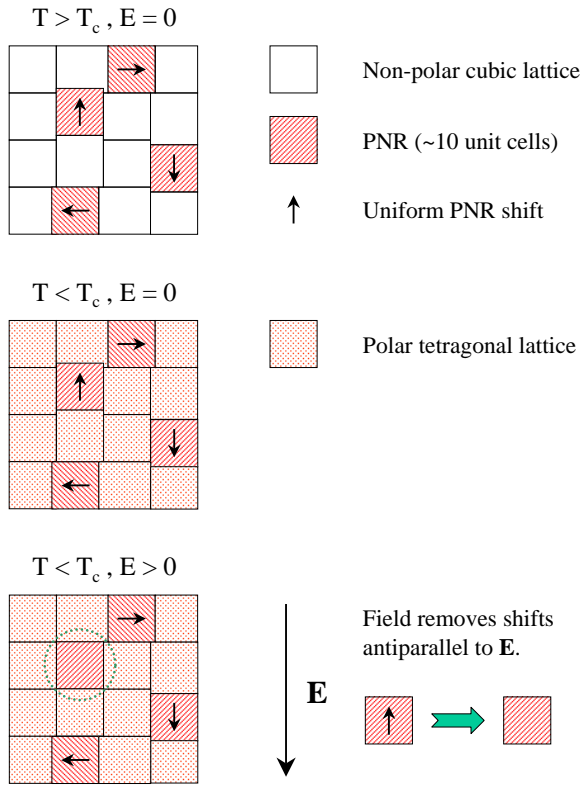}
\caption{\label{fig:6} Top panel:  Phase-shifted PNR embedded in a
non-polar cubic lattice for $T > T_c$.  Middle panel: Below $T_c$
the lattice achieves a polarization of its own, but the phase
shift remains.  Bottom panel:  An applied electric field removes
shifts along ${\bf E}$, while preserving those that are
orthogonal.}
\end{figure}
%
%

\section{Acknowledgments}

We would like to thank R.\ Erwin, Y.\ Fujii, K.\ Hirota, C.\
Stock, and S.\ Wakimoto, for stimulating discussions, and P.\ W.\
Rehrig for supplying the high quality PZN-8\%PT single crystal.
We acknowledge financial support from the U.\ S.\ Dept.\ of Energy
under contract No.\ DE-AC02-98CH10886.  We also acknowledge the
U.S. Dept. of Commerce, NIST Center for Neutron Research, for
providing the neutron scattering facilities used in this study.

\end{document}